     \documentstyle[12pt,epsfig]{article}   
     \newlength{\dinwidth}                       
     \newlength{\dinmargin}                      
\def\lsim{\mathrel{\rlap{\lower4pt\hbox{\hskip1pt$\sim$}}
    \raise1pt\hbox{$<$}}}                
\def\gsim{\mathrel{\rlap{\lower4pt\hbox{\hskip1pt$\sim$}}
    \raise1pt\hbox{$>$}}}                
\begin{document}
\vspace*{10mm}
\centerline{\normalsize hep--ph/9907315\hfill}
\begin{center}  \begin{Large} \begin{bf}
\boldmath
Higher-Order Effects on Inelastic $J/\psi$ Photoproduction\footnote{To appear
in the proceedings of the workshop on {\it Monte Carlo Generators for HERA 
Physics}, DESY, Hamburg, 21 April 1998 -- 5 February 1999, edited by T. Doyle,
G. Grindhammer, G. Ingelman, and H. Jung.}\\
\unboldmath
  \end{bf}  \end{Large}
  \vspace*{5mm}
  \begin{large}
B.A. Kniehl
  \end{large}
\end{center}
II. Institut f\"ur Theoretische Physik, Universit\"at Hamburg,
Luruper Chaussee 149, Germany\\
\begin{quotation}
\noindent
{\bf Abstract:}
Approximately taking into account the higher-order effects due to 
multi\-ple-gluon initial-state radiation, we extract from the latest Tevatron
data of prompt $J/\psi$ hadroproduction the leading colour-octet matrix
elements within the nonrelati\-vistic-QCD (NRQCD) factorization formalism.
We find that the matrix elements which describe the formation of $J/\psi$ 
mesons from colour-octet $c\bar c$ pairs in the angular-momentum states
${}^{2S+1}\!L_J={}^1\!S_0$ and ${}^3\!P_J$, with $J=0,1,2$, which are 
responsible for the excess of the predicted cross section of inelastic
$J/\psi$ photoproduction over the existing HERA data at high values of the 
inelasticity variable $z$, are significantly reduced.
As a consequence, the Tevatron and HERA measurements of inclusive $J/\psi$
production are reconciled in the NRQCD framework.
\end{quotation}

\section{Introduction}

The cross section of inclusive $J/\psi$ hadroproduction measured in
$p\bar p$ collisions at the Fermilab Tevatron \cite{abe,cdf} turned out to be
more than one order of magnitude in excess of what used to be the best
theoretical prediction, based on the colour-singlet model (CSM).
As a solution to this puzzle, Bodwin, Braaten, and Lepage \cite{bod} proposed
the existence of so-called colour-octet processes to fill the gap.
The central idea is that $c\bar c$ pairs are produced at short distances in
colour-octet states and subsequently evolve into physical (colour-singlet)
charmonia by the nonperturbative emission of soft gluons.
The underlying theoretical framework is provided by nonrelativistic QCD
(NRQCD) endowed with a particular factorization hypothesis, which implies a
separation of short-distance coefficients, which are amenable to perturbative
QCD, from long-distance matrix elements, which must be extracted from
experiment.
This formalism involves a double expansion in the strong coupling constant
$\alpha_s$ and the relative velocity $v$ of the bound charm quarks, and takes
the complete structure of the charmonium Fock space into account.

In the case of inelastic $J/\psi$ photoproduction, NRQCD with
colour-octet matrix elements tuned \cite{cho} to fit the Tevatron data
\cite{abe} predicts \cite{cac} at leading order (LO) a distinct rise in
cross section as $z\to1$, where $z$ is the fraction of the photon energy
transferred to the $J/\psi$ meson in the proton rest frame, which is not
observed by the H1 \cite{aid} and ZEUS \cite{bre} collaborations at HERA.
This colour-octet charmonium anomaly has cast doubts on the validity of the
NRQCD factorization hypothesis \cite{bod}, which seems so indispensible to
interpret the Tevatron data in a meaningful way.

Here, we report on an attempt \cite{kni} to rescue the NRQCD approach by
approximately taking into account dominant higher-order (HO) QCD effects.
The basic idea is as follows.
The predicted excess over the HERA data at $z$ close to unity is chiefly
generated by colour-octet $c\bar c$ pairs in the states ${}^1\!S_0$,
${}^3\!P_0$, and ${}^3\!P_2$ \cite{cac}, where we use the spectroscopic
notation ${}^{2S+1}\!L_J$ to indicate the spin $S$, the orbital angular
momentum $L$, and the total angular momentum $J$.
On the other hand, in hadroproduction at the Tevatron, the contributions from
the colour-octet ${}^1\!S_0$ and ${}^3\!P_J$ states fall off much more
strongly with increasing transverse momentum ($p_T$) than the one due to the
colour-octet ${}^3\!S_1$ state \cite{cho}, which is greatly suppressed in the
quasi-elastic limit of photoproduction \cite{cac}.
Consequently, the nonperturbative matrix elements which are responsible for
the colour-octet charmonium crisis are essentially fixed by the Tevatron data
in the low-$p_T$ regime.
This is precisely where the LO approximation used in Ref.~\cite{cho} is
expected to become unreliable due to multiple-gluon radiation from the
initial and final states.
In Ref.~\cite{can}, this phenomenon was carefully analyzed in a Monte Carlo
framework and found to significantly increase the LO cross section.
In Ref.~\cite{kni}, fits to the latest prompt $J/\psi$ data taken by the
CDF collaboration \cite{cdf} at the Tevatron were performed incorporating this
information \cite{can} on the dominant HO QCD effects.
The resulting HO-improved NRQCD predictions for inelastic $J/\psi$
photoproduction at HERA do not overshoot the H1 \cite{aid} and ZEUS \cite{bre}
data any more.

\section{Theoretical input}

The underlying theoretical framework is explained in Ref.~\cite{kni}.
If $p_T$ is of order $M_{J/\psi}$ or below, we adopt the fusion picture, where
the $c\bar c$ bound state is formed within the primary hard-scattering process.
In the high-$p_T$ regime, we work in the fragmentation picture, where the
$c\bar c$ bound state is created from a single high-energy gluon, charm quark
or antiquark which is close to its mass shell.
We take the renormalization scale $\mu$ and the common factorization scale
$M_f$ to be $\mu=M_f=m_T$, where $m_T=\sqrt{4m_c^2+p_T^2}$ is the $J/\psi$
transverse mass.
We define the starting scale $\mu_0$ of the fragmentation functions (FF's) as
$\mu_0=2m_c=M_{J/\psi}$.
For our LO analysis, we choose CTEQ4L \cite{lai} and GRV-LO \cite{glu} as the
proton and photon PDF's, respectively, and evaluate $\alpha_s$ from the
one-loop formula with $\Lambda^{(4)}=236$~MeV \cite{lai}.
Whenever we include higher orders, we adopt the $\overline{\rm MS}$
renormalization and factorization scheme and employ CTEQ4M \cite{lai}, GRV-HO
\cite{glu}, and the two-loop formula for $\alpha_s$ with
$\Lambda_{\overline{\rm MS}}^{(4)}=296$~MeV \cite{lai}.

Unfortunately, not all ingredients which would be necessary for a fully
consistent NLO analysis are yet available.
In the case of fusion, the NLO corrections to the partonic cross sections are
only known for direct photoproduction in the CSM \cite{kra}.
Furthermore, in the case of fragmentation, the NLO corrections to the FF's
at the initial scale $\mu_0$ are still unknown.
In the case of direct $J/\psi$ photoproduction under typical HERA conditions,
the QCD correction factor $K$ to the inclusive cross section in the CSM was
found \cite{kra} to be as low as 1.2 in the inelastic regime $z\lsim0.9$.
It is plausible that the $K$ factors for the color-octet and resolved-photon
processes should be modest, too.
However, the situation should be very different for inclusive $J/\psi$
hadroproduction at the Tevatron, especially in the low-$p_T$ range, where one
expects substantial HO QCD effects due to multiple-gluon radiation.
Such effects were estimated for the fusion mechanism in Ref.~\cite{can} by
means of the Monte Carlo event generator PYTHIA \cite{sjo} after implementing
therein the relevant colour-octet processes, and they were indeed found to be
very sizeable.
The impact of these effects on the fit to the latest CDF data \cite{cdf} is
demonstrated in Table~\ref{t1}.

\begin{table}
\begin{center}
\caption{Values of the $J/\psi$ matrix elements resulting from the
LO and HO-improved fits to the CDF data \protect\cite{cdf}.
Here, $M_r^{J/\psi}=
\left\langle{\cal O}^{J/\psi}[\,\underline{8},{}^1\!S_0]\right\rangle
+(r/m_c^2)
\left\langle{\cal O}^{J/\psi}[\,\underline{8},{}^3\!P_0]\right\rangle$.
The first 10 (last 2) of the 11 data points are described in the fusion
(fragmentation) picture.}
\label{t1}
\medskip
\begin{tabular}{|c|c|c|}
\hline\hline
 & LO & HO \\
\hline
$\left\langle{\cal O}^{J/\psi}[\,\underline{1},{}^3\!S_1]\right\rangle$ &
$(7.63\pm0.54)\cdot10^{-1}$~GeV$^3$ & $(1.30\pm0.09)$~GeV$^3$ \\
$\left\langle{\cal O}^{J/\psi}[\,\underline{8},{}^3\!S_1]\right\rangle$ &
$(3.94\pm0.63)\cdot10^{-3}$~GeV$^3$ & $(2.73\pm0.45)\cdot10^{-3}$~GeV$^3$ \\
$M_r^{J/\psi}$ &
$(6.52\pm0.67)\cdot10^{-2}$~GeV$^3$ & $(5.72\pm1.84)\cdot10^{-3}$~GeV$^3$ \\
$r$ & 3.47 & 3.54 \\
$\chi_{\rm DF}^2$ & 7.49/12 & 3.96/12 \\
\hline\hline
\end{tabular}
\end{center}
\end{table}

\section{Predictions for charmonium photoproduction}

We now explore the phenomenological consequences of this HO improvement for
inclusive $J/\psi$ photoproduction in $ep$ collisions at HERA, with beam
energies $E_e=27.5$~GeV and $E_p=820$~GeV in the laboratory frame, assuming
the maximum photon virtuality to be $Q_{\rm max}^2=4$~GeV$^2$.
As in the H1 \cite{aid} and ZEUS \cite{bre} publications, we convert the $ep$
cross sections to averaged $\gamma p$ cross sections by dividing out the
photon-flux factor.
The contribution due to $\psi^\prime$ mesons with subsequent decay into
$J/\psi$ mesons is approximately taken into account by multiplying the
theoretical predictions by an overall factor of 1.15.
The data are mostly concentrated in the low-$p_T$ range, where the fusion
picture should be valid.

In Fig.~1, we compare our LO and HO-improved predictions for the
$p_T^2$, $z$, and $y$ distributions with the H1 \cite{aid} and ZEUS \cite{bre}
data.
Here, $y$ is the $J/\psi$ rapidity in the laboratory frame, which is taken to 
be positive in the proton flight direction.
The circumstance that
$\left\langle{\cal O}^{J/\psi}[\,\underline{8},{}^1\!S_0]\right\rangle$ and
$\left\langle{\cal O}^{J/\psi}[\,\underline{8},{}^3\!P_0]\right\rangle$
are not separately fixed by the fit to the CDF data \cite{cdf} induces some
uncertainty 
in the colour-octet contributions to the cross sections of direct and resolved
photoproduction and thus also in the total cross section.
This uncertainty is encompassed by the results for
$\left\langle{\cal O}^{J/\psi}[\,\underline{8},{}^1\!S_0]\right\rangle=
M_r^{J/\psi}$ and
$\left\langle{\cal O}^{J/\psi}[\,\underline{8},{}^3\!P_0]\right\rangle=0$ and
those for
$\left\langle{\cal O}^{J/\psi}[\,\underline{8},{}^1\!S_0]\right\rangle=0$ and
$\left\langle{\cal O}^{J/\psi}[\,\underline{8},{}^3\!P_0]\right\rangle
=(m_c^2/r)M_r^{J/\psi}$, which are actually shown in Fig.~1.
We observe that, at LO, the colour-octet contribution of direct
photoproduction is dominant for $z\gsim0.5$.
Thus, it also makes up the bulk of the $p_T^2$ and $y$ distributions,
which are integrated over $0.4<z<0.9$.
This contribution is responsible for the significant excess of the LO
predictions over the experimental results for $d\sigma/dp_T^2$ and
$d\sigma/dz$ at low $p_T$ and high $z$, respectively.
On the other hand, the HO-improved predictions tend to undershoot the data
leaving room for a substantial $K$ factor due to the missing NLO corrections
to the partonic cross sections.
Now, the colour-singlet contribution of direct photoproduction, which is well
under theoretical control \cite{kra}, is by far dominant, except in the
corners of phase space, at $z\lsim0.15$ and $z\gsim0.85$, where the colour-octet
contributions of resolved and direct photoproduction, respectively, take over.
Of course, we should also bear in mind that the predictions shown in
Fig.~1 still suffer from considerable theoretical
uncertainties related to the choice of the scales $\mu$ and $M_f$, the PDF's,
and other input parameters such as $m_c$ and $\Lambda^{(4)}$ \cite{kra}.
From these observations, we conclude that it is premature at this point to
speak about a discrepancy between the Tevatron \cite{abe,cdf} and HERA
\cite{aid,bre} data of inclusive $J/\psi$ production within the framework of
NRQCD \cite{bod}.

\begin{figure}[ht]
\label{f1}
\begin{tabular}{ll}
\parbox{7.5cm}{
\epsfig{file=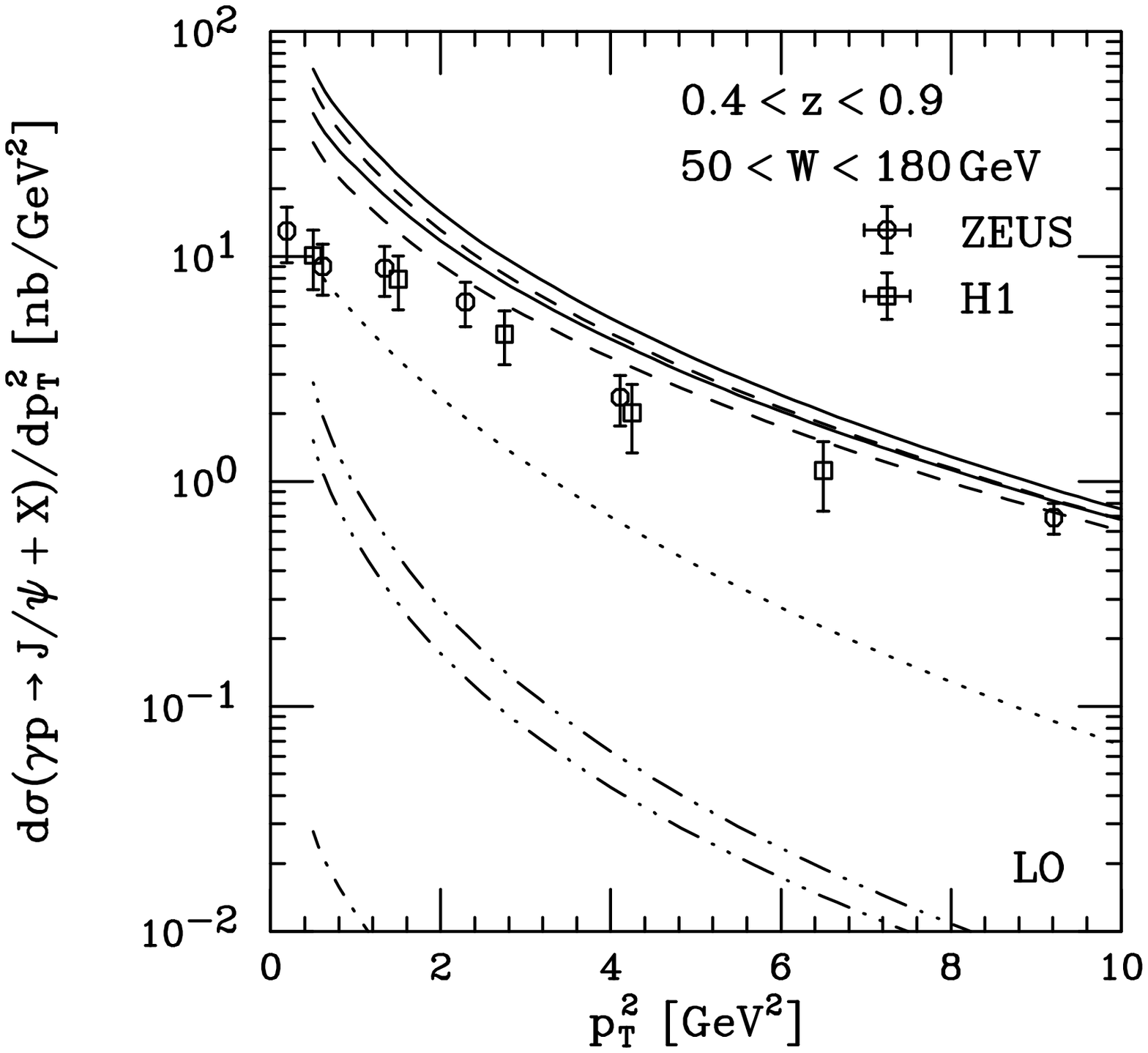,width=7.5cm}
}
&
\parbox{7.5cm}{
\epsfig{file=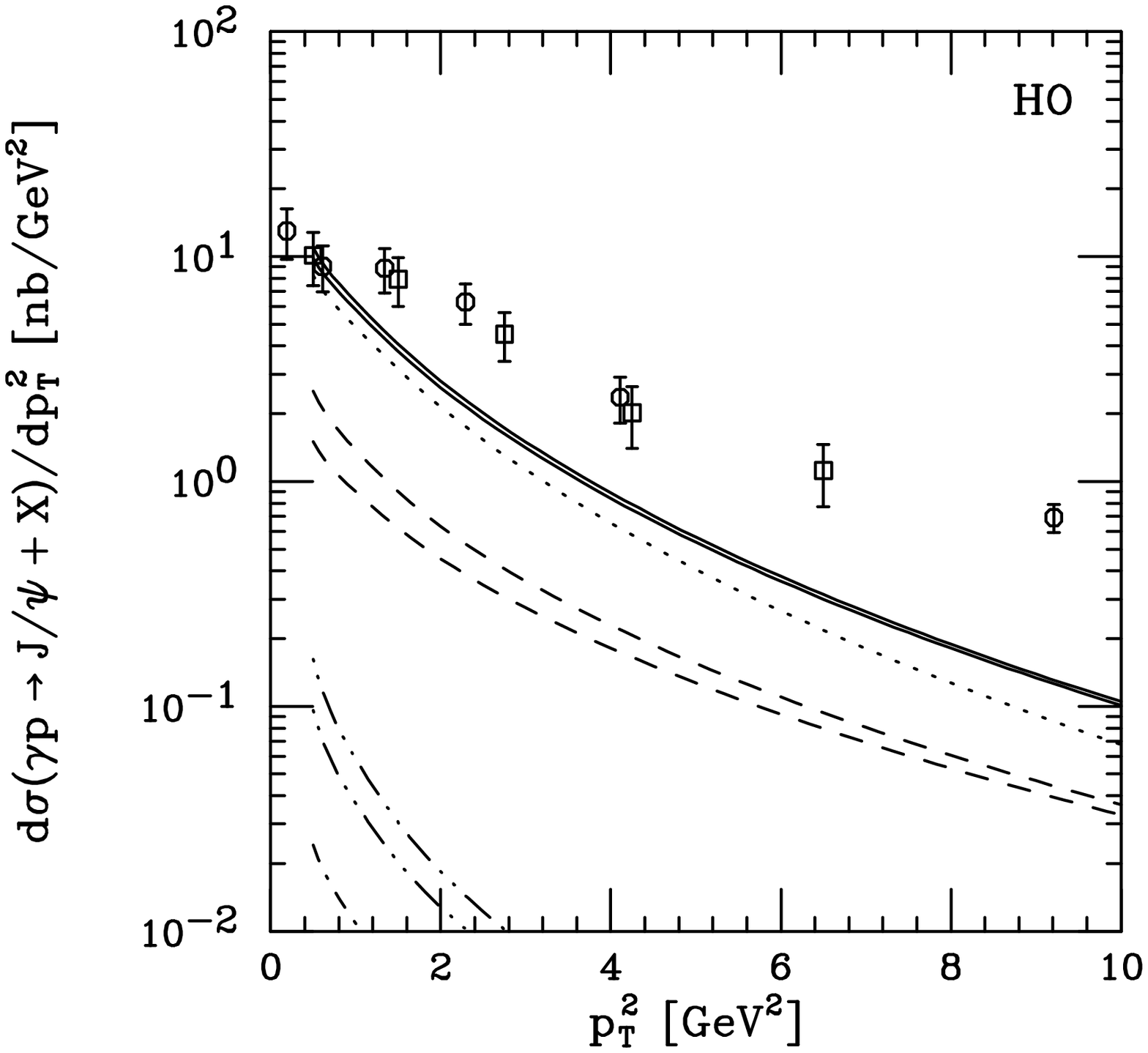,width=7.5cm}
}
\\
\vspace*{1mm}
\parbox{7.5cm}{
\epsfig{file=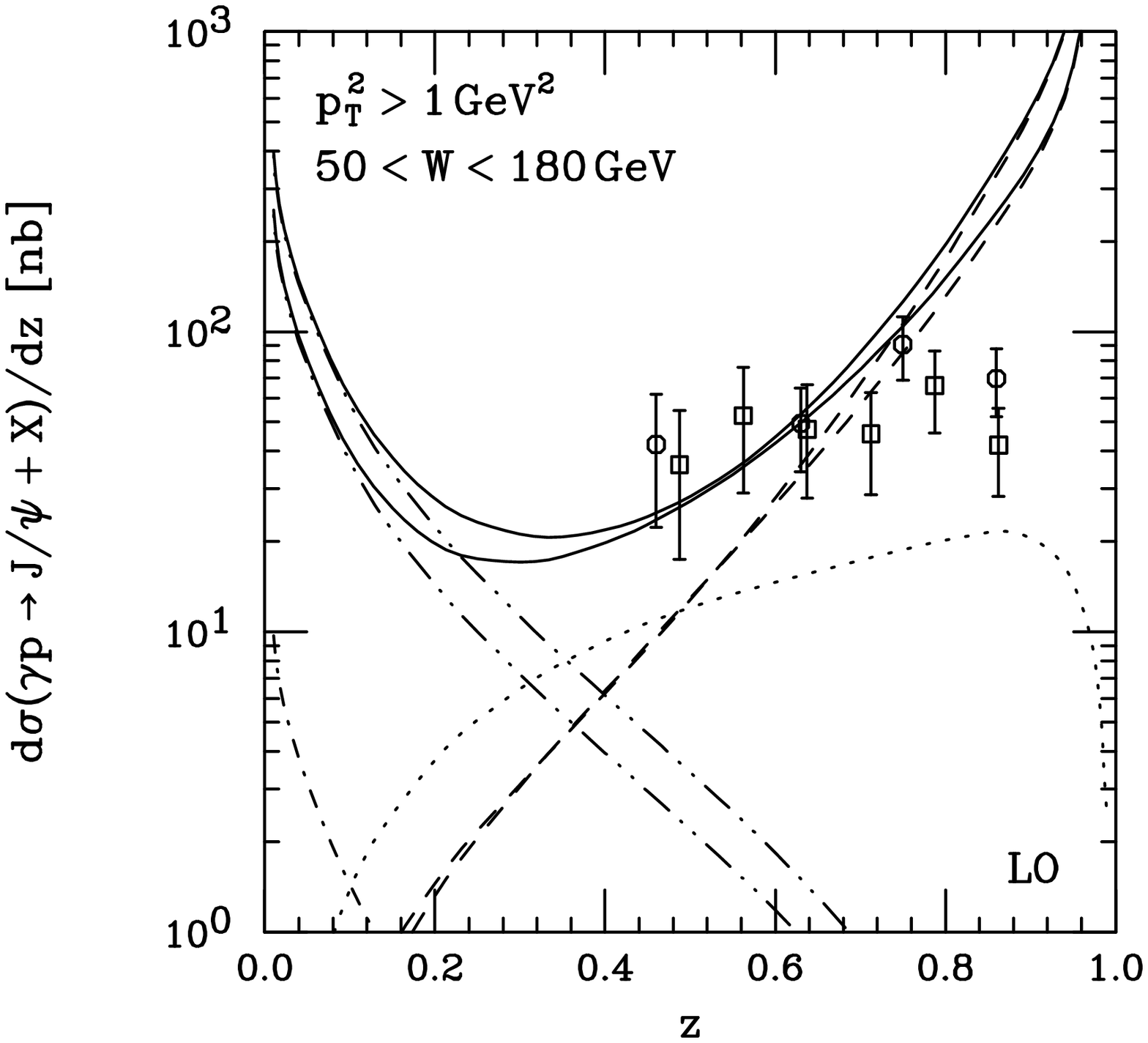,width=7.5cm}
}
&
\parbox{7.5cm}{
\epsfig{file=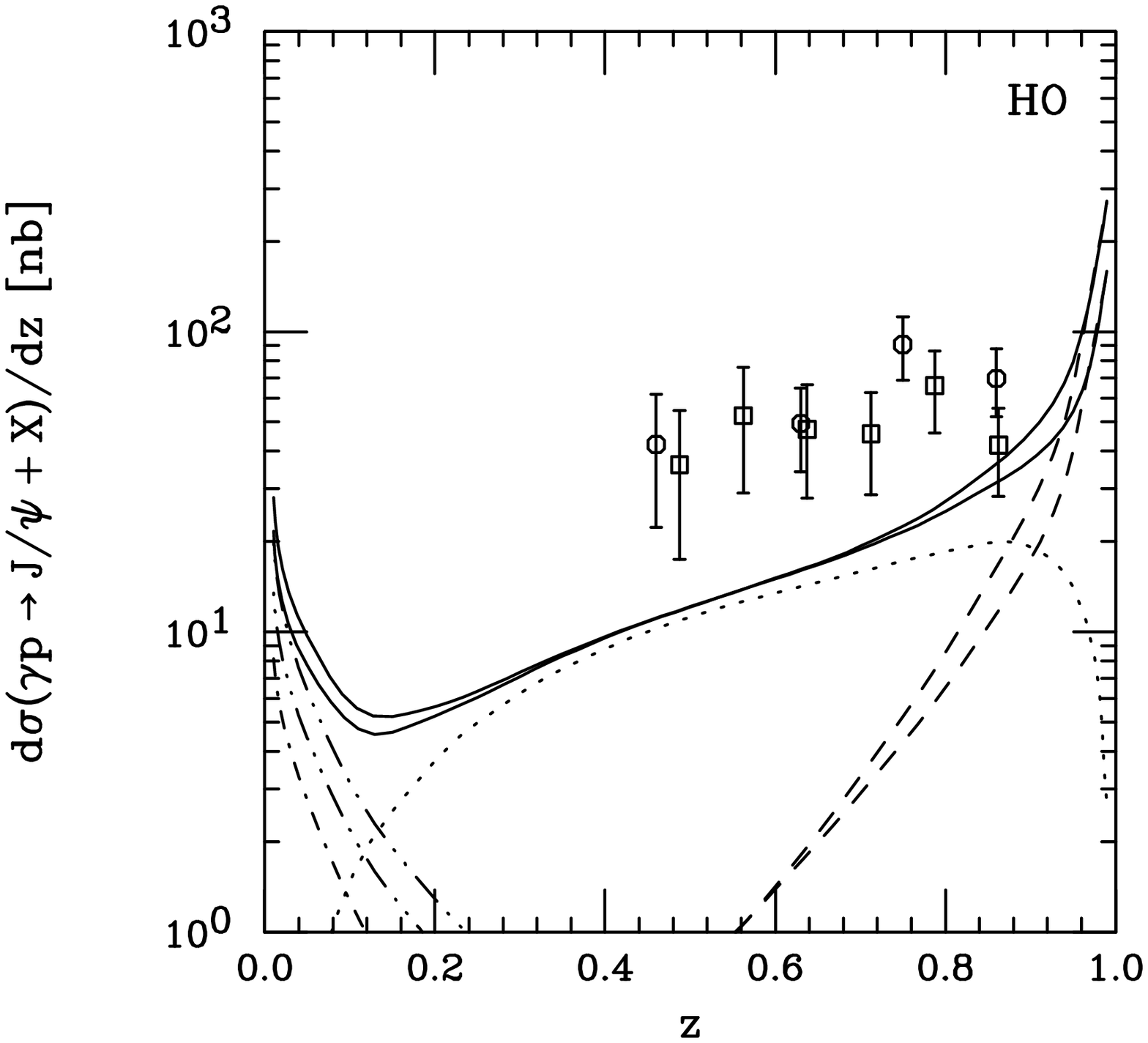,width=7.5cm}
}
\\
\vspace*{1mm}
\parbox{7.5cm}{
\epsfig{file=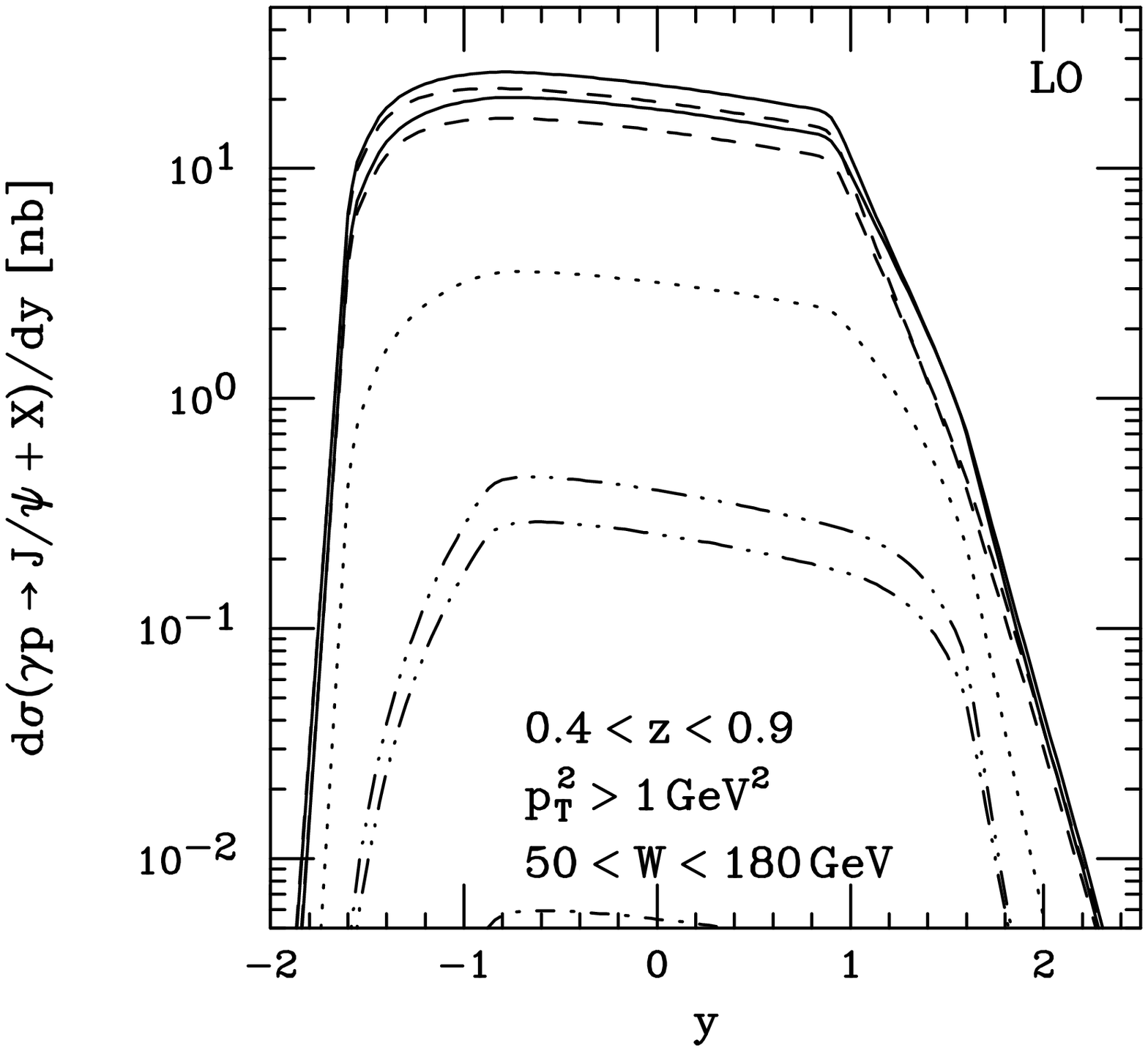,width=7.5cm}
}
&
\parbox{7.5cm}{
\epsfig{file=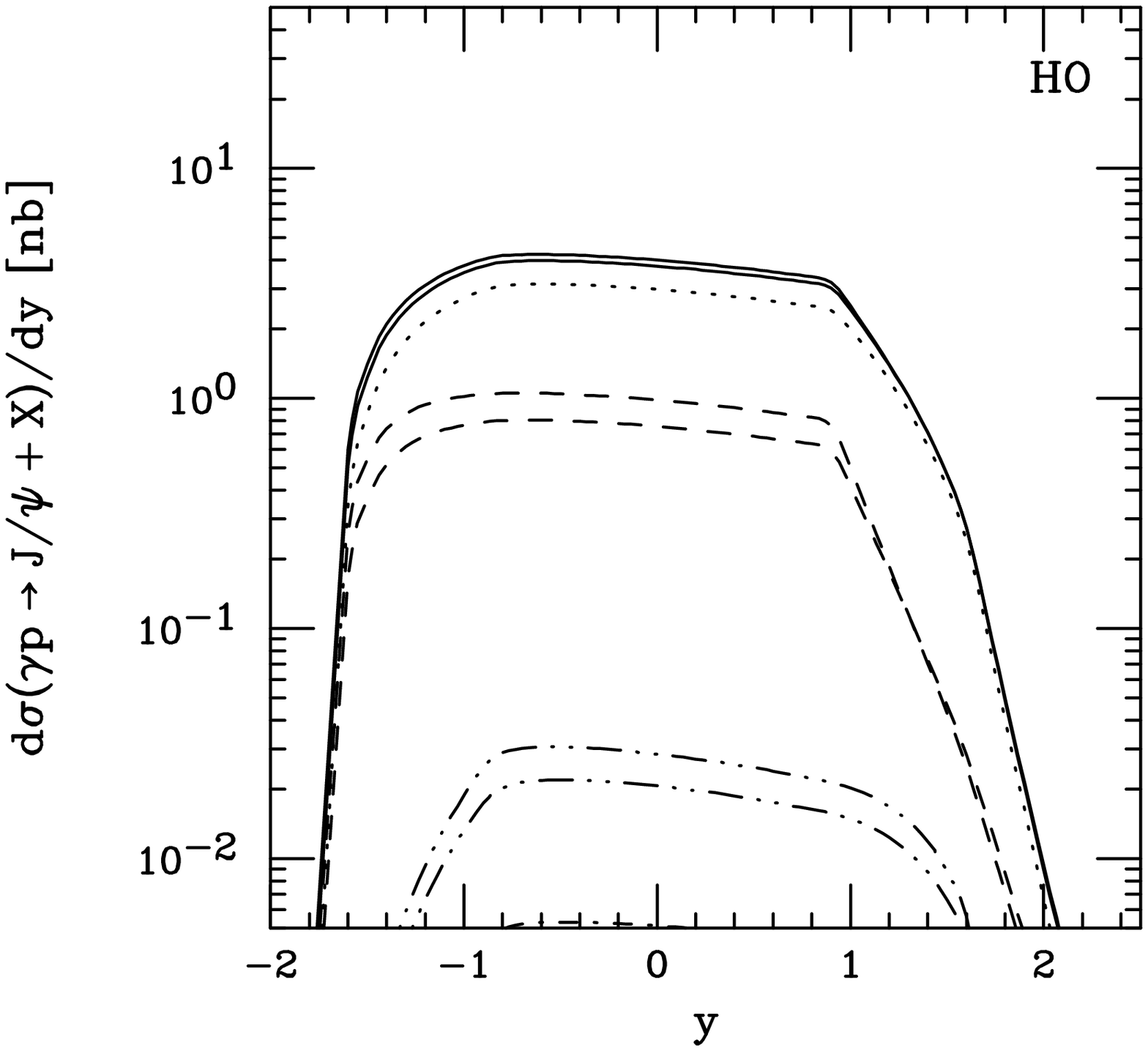,width=7.5cm}
}
\end{tabular}
\vspace*{2mm}
\caption[junk]{{\it
The H1 \protect\cite{aid} and ZEUS \protect\cite{bre} data on inelastic
$J/\psi$ photoproduction are compared with the LO and HO-improved predictions.
The total results (solid lines) are built up by the
direct-photon colour-singlet (dotted lines),
direct-photon colour-octet (dashed lines),
resolved-photon colour-singlet (dot-dashed lines), and
resolved-photon colour-octet (dot-dot-dashed lines) contributions.
}}
\end{figure}

\section{Conclusions}

We determined the $J/\psi$ colour-octet matrix elements which appear in the
NRQCD expansion \cite{bod} at leading order in $v$ by fitting the latest
Tevatron data of prompt $J/\psi$ hadroproduction \cite{cdf}.
We found that the result for the linear combination $M_r^{J/\psi}$ of
$\left\langle{\cal O}^{J/\psi}[\,\underline{8},{}^1\!S_0]\right\rangle$ and
$\left\langle{\cal O}^{J/\psi}[\,\underline{8},{}^3\!P_0]\right\rangle$ is
substantially reduced if the HO QCD effects due to the multiple emission of
gluons, which had been estimated by Monte Carlo techniques \cite{can}, are
taken into account.
As an important consequence, the intriguing excess of the LO NRQCD prediction
for inelastic $J/\psi$ photoproduction at $z$ close to unity \cite{cac}
over the HERA measurements \cite{aid,bre} disappears.
We assess this finding as an indication that it is premature to proclaim an
experimental falsification of the NRQCD framework on the basis of the HERA 
data.
Although we believe that our analysis captures the main trend of the HO
improvement, we stress that it is still at an exploratory level, since a
number of ingredients which would be necessary for a fully consistent NLO
treatment of inclusive $J/\psi$ hadroproduction and photoproduction are still
missing.

\medskip

\noindent
{\it Achnowledgements.} The author thanks Gustav Kramer for his collaboration
on this work.

\end{document}